\documentclass{article}

\usepackage{amsmath}
\usepackage{pst-plot} 

\usepackage{amsthm}
\numberwithin{equation}{section}
\newtheorem{theorem}{Theorem}
\newtheorem*{acknowledgments}{Acknowledgments}
\theoremstyle{definition}
\newtheorem*{example}{Example}

\newcommand{\abs}[1]{\left\vert{#1}\right\vert}
\newcommand{\sgn}{\mathop{\mathrm{sgn}}\nolimits}
\newcommand{\R}{\mathbf{R}}
\newcommand{\tcoll}{t_{\mathrm{coll}}}

\begin{document}

\title{Continuous and discontinuous piecewise linear solutions of the
  linearly forced inviscid Burgers equation}

\author{Hans Lundmark\thanks{Department of Mathematics, Link{\"o}ping University, SE-581 83 Link{\"o}ping, SWEDEN; halun@mai.liu.se} \and Jacek Szmigielski\thanks{Department of Mathematics and Statistics, University of Saskatchewan, 106 Wiggins Road, Saskatoon, Saskatchewan, S7N 5E6, CANADA; szmigiel@math.usask.ca}}

\date{March 31, 2008}

\maketitle

\begin{abstract}
  \noindent
  We study a class of piecewise linear solutions to the inviscid Burgers
  equation driven by a linear forcing term.
  Inspired by the analogy with peakons, we think of these solutions as being
  made up of solitons situated at the breakpoints.
  We derive and solve ODEs governing the soliton dynamics,
  first for continuous solutions, and then for more general shock
  wave solutions with discontinuities.
  We show that triple collisions of solitons cannot take place
  for continuous solutions, but give an example of a triple collision in
  the presence of a shock.
\end{abstract}

\section{Introduction}

The subject of this paper is piecewise linear solutions of the PDE
\begin{equation}
  \label{eq:dB}
  (u_t+u u_x)_{xx} = 0,
\end{equation}
which we earlier \cite{kls} have called the \emph{derivative Burgers equation}.
This name refers of course to the well-known Burgers equation $u_t+u u_x=\nu u_{xx}$ and its special case
the inviscid Burgers equation $u_t+u u_x=0$,
which is the prototype equation for studying
shock wave solutions of hyperbolic conservation laws.
In some applications
one considers also \emph{forced} Burgers equations with terms of the form
$F(x,t)$ on the right-hand side, often written as $F=-\partial V/\partial x$
with a potential~$V$.
Since equation \eqref{eq:dB} is equivalent to
$u_t+u u_x = A(t)x+B(t)$,
it is perhaps more appropriate to talk about it as a
\emph{forced inviscid Burgers equation with linear force} (or quadratic potential).
Moreover, the latter equation can be rewritten as 
\begin{equation} \label{eq:fB}
u_t+\frac{1}{2}(u^2)_x=A(t)x +B(t),
\end{equation}
which makes sense for a much larger class of functions than
just $u\in C^1(\R)$. For example, if
$u\in L^2_{\mathrm{loc}}(\R)$ we can interpret
\eqref{eq:fB} to hold in the sense of distributions.
One could work with distributions acting on test functions
$\psi(x,t) \in \mathcal{D}(\R^2)$,
but the following simpler interpretation is sufficient for our
purposes here: we view $u(x,t)$ as a mapping that takes
a real number $t$ to a function $u(\cdot,t) \in L^2_{\mathrm{loc}}(\R)$
which we can identify with a distribution in $\mathcal{D}'(\R)$.
The derivative with respect to $x$ is then the distributional derivative
defined by its action on a test function $\psi(x)\in\mathcal{D}(\R)$ in
the usual way, $\langle u_x, \psi \rangle = - \langle u, \psi_x \rangle$,
while the derivative with respect to $t$ is the limit of a difference quotient.
If equation \eqref{eq:fB} is satisfied in $\mathcal{D}'(\R)$ for each $t$,
then we then say that it holds in a weak sense and that $u$ is its weak solution.

We were led to the Burgers equation by our
previous work on \emph{peakon} and \emph{shockpeakon} solution of the
Degasperis--Procesi (DP) equation
\begin{equation}
  \label{eq:DP}
  u_t-u_{txx} + 4u u_x = 3 u_x u_{xx} + u u_{xxx},
\end{equation}
an integrable wave equation discovered a few years ago \cite{dp,dhh1}.
Indeed, the problems treated in this paper are to some extent
``toy problems'', but we hope that they might provide some guidance and
intuition for the future study of the DP equation.

Equation \eqref{eq:dB} can be obtained formally from the DP equation
by substituting $x\mapsto \varepsilon x$, $t\mapsto \varepsilon t$,
and then letting $\varepsilon\to 0$.
This ``high-frequency limit'' is a natural thing to try on the
DP equation, since it is the same procedure that takes
the celebrated integrable Camassa--Holm (CH) shallow water equation \cite{ch},
\begin{equation}
  \label{eq:CH}
  u_t-u_{txx} + 3u u_x = 2 u_x u_{xx} + u u_{xxx},
\end{equation}
to the Hunter--Saxton (HS) equation for nematic liquid crystals
\cite{hunter-saxton,hz1},
\begin{equation}
  \label{eq:HS}
  (u_t+u u_x)_{xx} = u_x u_{xx}.
\end{equation}

The CH and DP equations both admit \emph{peakon} solutions,
which are multisoliton solutions of the form
\begin{equation}
  \label{eq:peakon}
  u(x,t) = \sum_{k=1}^n m_k(t)\,\exp\bigl(-\abs{x-x_k(t)} \bigr),
\end{equation}
where the functions $x_k(t)$ and $m_k(t)$
(positions and momenta of the individual peak-shaped solitons)
are required to satisfy a certain system of $2n$ ODEs in order for
$u(x,t)$ to satisfy the PDE in a weak sense.
In shorthand notation these ODEs are $\dot{x}_k = u(x_k)$,
$\dot{m}_k = -(b-1) u_x(x_k)$, where $b=2$ for the CH equation
and $b=3$ for the DP equation.
One can think of this as an integrable mechanical system of $n$ particles
on the real line, simililar to, for example, the open Toda lattice.
If follows from the rapid decay of $e^{-\abs{x}}$
that $\dot{x}_k = u(x_k) \approx m_k$
when all distances $\abs{x_i-x_j}$ are large,
so it agrees with intuition to regard $m_k$ as the momentum of the
$k$th particle.
Asympotically (when $t\to\pm\infty$) the particles will spread apart,
each moving with its own (nearly) constant velocity which is nonzero
and distinct from the other particles' velocities.
The latter is a highly nontrivial
fact for the DP equation \cite[Theorem~2.4]{ls-imrp}.

There is an analogous class of solutions of the HS equation
and the forced Burgers equation \eqref{eq:fB},
namely the piecewise linear solutions
\begin{equation}
  \label{eq:u-ansatz}
  u(x,t) = \sum_{k=1}^n m_k(t) \abs{x-x_k(t)}.
\end{equation}
In the shorthand notation used above, the governing ODEs take exactly the
same form again:
$\dot{x}_k = u(x_k)$, $\dot{m}_k = -(b-1) u_x(x_k)$,
where $b=2$ for the HS equation and $b=3$ for the forced Burgers equation.
However, for peakons the term $m_k e^{-\abs{x_k-x_k}}$
usually dominates the other terms in the equation $\dot{x}_k = u(x_k)$,
while here we instead have the term
$m_k \abs{x_k-x_k}$ which is zero while all \emph{other} terms are large.
Thus, in contrast to peakons where the interaction is strongly localized,
these piecewise linear solitons influence each other more strongly the more
separated they are.
Although it is a bit hard to develop a useful intuition about
these ODEs as a ``mechanical'' system
(perhaps one can think of some kind of expanding gas with long-range correlations),
the analogy with peakons still
makes it natural to think of the piecewise linear solutions as
being composed of some kind of solitons situated at the
breakpoints~$x_k$.
(But we have not been able
to make sense of the idea that the piecewise linear solutions are
somehow high-frequency limits of peakons).

In all four cases mentioned above,
the ODEs governing the soliton dynamics can be explicitly solved using
inverse spectral methods
\cite{bss-moment,ls-invprob,ls-imrp,bss-HS,kls,bressan-constantin-CH}.
In the forced Burgers case the ODEs are also easily solved
directly by elementary methods, as we will see.

In the Degasperis--Procesi equation 
(but not in the Camassa--Holm equation)
there also appears a more complicated phenomenon,
namely discontinuous solutions of the form
\begin{equation}
  \label{eq:shockpeakon}
  u(x,t) = \sum_{k=1}^n \biggl( m_k(t)
  - s_k(t)\,\sgn\bigl( x - x_k(t) \bigr) \biggr) \exp\bigl(- \abs{x-x_k(t)} \bigr).
\end{equation}
Such \emph{shockpeakons} \cite{lundmark-shockpeakons}
are governed by $3n$ ODEs for positions $x_k$,
momenta $m_k$, and shock strengths~$s_k$.
Even if one starts with the usual peakon ansatz \eqref{eq:peakon},
shock solutions of the form \eqref{eq:shockpeakon} can form
after finite time when a peakon with $m_k > 0$ collides with
an \emph{antipeakon} with $m_{k+1}<0$ moving in the opposite direction.
(In the CH equation, such collisions give rise to ``zero-strength shocks''
where $u_x$ momentarily blows up but $u$ remains continuous,
still being of the form \eqref{eq:peakon} after the collision \cite{bss-moment},
and a similar thing occurs for the HS equation \cite{hz1}.)
The shockpeakon ODEs have so far only been solved in the trivial case $n=1$
and in a very particular subcase when $n=2$.
The problem is that the Lax pair for the DP equation, which was crucial
for deriving the peakon solution formulas,
does not make sense for the weak formulation of the DP equation that is
used when working with discontinuous solutions.

The forced Burgers equation \eqref{eq:fB} admits an analogous class of solutions,
given by the discontinuous piecewise linear ansatz
\begin{equation}
  \label{eq:u-shock-ansatz}
  u(x,t) = \sum_{k=1}^n \biggl( m_k(t) \abs{x-x_k(t)} - s_k(t) \sgn\bigl( x - x_k(t) \bigr) \biggr).
\end{equation}
Such solutions with shocks can form after finite time, even if the initial
profile is continuous.
Unlike the Degasperis--Procesi case, it turns out here that
the extra generality of having jumps in $u$
can be handled without problems.

The outline of the paper is simple: we derive and solve the ODEs govering
piecewise linear solutions of the forced Burgers equation \eqref{eq:fB},
first in the simpler case \eqref{eq:u-ansatz} of continuous solutions
(using elementary methods and, for comparison, inverse spectral methods),
then in the general case \eqref{eq:u-shock-ansatz} of discontinuous solutions
(by reduction to the previous case).
We conclude with a few examples.

\section{Continuous piecewise linear solutions}

\begin{theorem}
  \label{thm:ODEs}
  The continuous piecewise linear ansatz \eqref{eq:u-ansatz},
  $u=\sum m_k \abs{x-x_k}$,
  is a weak solution to the linearly forced inviscid Burgers equation 
  \eqref{eq:fB} if and only if
  \begin{equation}
    \label{eq:ODEs}
     \dot{x}_k =\sum_{i=1}^n m_i \abs{ x_k-x_i},
    \qquad
    \dot{m}_k = 2 m_k \sum_{i=1}^n m_i \sgn(x_i-x_k),
  \end{equation}
  for $k=1,\ldots,n$.
  For this class of solutions, equation \eqref{eq:fB} takes the form
  \begin{equation}
    \label{eq:fB-with-ansatz}
    u_t+\frac{1}{2}(u^2)_x = M^2 x - MM_+,
  \end{equation}
  where $M=\sum_{k=1}^n m_k$ and $M_+=\sum_{k=1}^n m_k x_k$
  are constants of motion.
\end{theorem}

\begin{proof}
  This is a special case (all $s_k=0$) of
  Theorem~\ref{thm:shock-ODEs} which is proved later.
\end{proof}

One can assume that all $m_k\neq 0$,
since it follows from \eqref{eq:ODEs} that any vanishing $m_k$ remains
identically zero.
If we think of $x_k$ and $m_k$ as positions and masses of particles on a line,
then the total mass $M$ and the center of mass $M_+/M$ (if $M \ne 0$)
are conserved.
Note that when $M=0$ we have the unforced Burgers equation.
There are some additional constants of motion $M_2, \dots, M_n$ that come
together with $M_1=M$ from the Lax pair presented in the next section,
but we will not need them here \cite{kls}.

The presence of absolute values and the sign function in
\eqref{eq:ODEs} naturally divides the position space $\R^n$ into sectors.
More precisely, to any permutation
$\sigma=\sigma_1 \sigma_2 \dots \sigma_n$ of the numbers
$\left\{ 1, 2, \dots, n \right\}$ one can assign the sector
$X_{\sigma}=\{ (x_1,x_2,\dots,x_n)\in\R^n \, |
\, x_{\sigma_1} < x_{\sigma_2} < \cdots < x_{\sigma_n} \}$.
We will concentrate on the sector $X_e$
corresponding to the identity permutation $e=$\mbox{$12\dots n$},
since there is no loss of generality in assuming that the initial
positions $x_k(0)$ are sorted in increasing order:
\begin{equation}
  \label{eq:identitysector}
  X_e = \{ (x_1,x_2,\dots,x_n)\in\R^n \, |
  \, x_1 < x_2 < \cdots < x_n \}.
\end{equation}
For positions in $X_e$ the ODEs \eqref{eq:ODEs} take the form
\begin{equation}
  \label{eq:ODEs-identitysector}
     \dot{x}_k =\sum_{i=1}^n m_i (x_k-x_i) \sgn(k-i),
    \qquad
    \dot{m}_k = 2 m_k \sum_{i=1}^n m_i \sgn(i-k).
\end{equation}
The following theorem solves this system completely.

\begin{theorem}
  \label{thm:ODE-solution}
  Given any initial data $\left\{ x_k(0), m_k(0) \right\}_{k=1}^n$
  (with the $x_k(0)$'s ordered or not),
  the solution of the ODEs \eqref{eq:ODEs-identitysector} is
  given by the formulas below, where $M=\sum m_k$ and
  $M_+ = \sum m_k x_k$ as before,
  and where the empty sums $\sum_1^0$ and $\sum_{n+1}^n$ in $F_0$ and $F_n$
  are to be interpreted as zero (so that $F_0(t)=e^{-Mt}$ and
  $F_n(t)=e^{Mt}$).
  \begin{itemize}
  \item When $M \ne 0$ the solution of \eqref{eq:ODEs-identitysector} is
    \begin{equation}
      \label{eq:ODEsolution}
      \begin{split}
        x_k(t) & = \frac{M_+}{M} + 
        \frac{e^{Mt}}{M} \left( \displaystyle \sum_{j<k} \bigl( x_k(0)-x_j(0) \bigr) m_j(0) \right)
        \\
        & \quad + \frac{e^{-Mt}}{M} \left( \displaystyle \sum_{j>k} \bigl( x_k(0)-x_j(0) \bigr) m_j(0) \right),
        \\
        m_k(t) &= \frac{m_k(0)}{F_{k-1}(t) F_k(t)},
      \end{split}
    \end{equation}
    for $k=1,\dots,n$, where
    \begin{equation}
      \label{eq:Fk}
      F_k(t) =\displaystyle
      \frac{e^{Mt}}{M} \left( \displaystyle \sum_{j=1}^k m_j(0) \right)
      + \frac{e^{-Mt}}{M} \left( \displaystyle \sum_{j=k+1}^n m_j(0) \right).
    \end{equation}
  \item When $M=0$ the solution of \eqref{eq:ODEs-identitysector} is
    \begin{equation}
      \begin{split}
        x_k(t) &= x_k(0) +
        t \left( \displaystyle \sum_{j<k} \bigl( x_k(0)-x_j(0) \bigr) m_j(0)
          - \displaystyle \sum_{j>k} \bigl( x_k(0)-x_j(0) \bigr) m_j(0) \right),
        \\
        m_k(t) &= \frac{m_k(0)}{F_{k-1}(t) F_k(t)}, 
      \end{split}
    \end{equation}
    for $k=1,\dots,n$, where
    \begin{equation}
      \label{eq:Fk-linear}
      F_k(t) =
      1 + t \, \left( \displaystyle\sum_{j=1}^k m_j(0) - \sum_{j=k+1}^n m_j(0) \right).
    \end{equation}
  \item 
    Letting $l_k = x_{k+1} - x_k$ for $k=1,\ldots,n-1$, we have in both cases
    \begin{equation}
      \label{eq:ODEsolution-lk}
      l_k(t) = l_k(0) F_k(t).
    \end{equation}
  \end{itemize}
\end{theorem}

The proof is presented at the end of this section.
As an immediate corollary we obtain information about the
original ODEs \eqref{eq:ODEs}.

\begin{theorem}
  Given initial data $\left\{ x_k(0), m_k(0) \right\}_{k=1}^n$
  to the ODEs \eqref{eq:ODEs}
  such that $x_1(0) < x_2(0) < \dots < x_n(0)$
  (that is, with the positions in the sector $X_e$ of $\R^n$),
  the solution is given locally (around $t=0$) by the formulas of
  Theorem~\ref{thm:ODE-solution}, and this solution is valid
  as long as the positions $x_k(t)$ remain in $X_e$.
\end{theorem}

A local solution that starts in $X_e$ hits the boundary of 
$X_e$ whenever $x_k = x_{k+1}$ for at least one~$k$,
an event which we refer to as a \emph{collision}.
It is clear from \eqref{eq:ODEsolution-lk} that a collision occurs
when some $F_k$ becomes zero,
at which time $m_k$ and $m_{k+1}$ blow up.
The local solution is valid up until the time of the first collision.
In general a shock will then form,
and the continuous ansatz \eqref{eq:u-ansatz} will not
be able to describe the solution beyond the point of collision.
We will return to this in the section about discontinuous solutions.

If all $m_k(0)$'s have the same sign, then \eqref{eq:Fk} shows
that there are no collisions, so the solution is global.
In the case when all are positive,
the asymptotic behaviour of this global solution
as $t\to+\infty$ is that $x_1 \to M_+ / M$ and $m_1 \to M$,
while $x_k \to+\infty$ and $m_k\to 0$ for all $k>1$.
When the $m_k(0)$'s have mixed signs, collisions may or may not
occur for $t>0$. For example, in the case $n=2$ a collision takes place
when $F_1(t) = (m_1(0) e^{Mt} + m_2(0) e^{-Mt})/M$ becomes zero,
which happens when $m_2(0) / m_1(0) < 0$ and
$t = (2M)^{-1} \ln\abs{m_2(0) / m_1(0)}$.
Consideration of cases shows that this value of $t$ is positive
iff $m_1(0) < 0 < m_2(0)$.

The event when $l_{k-1}=l_k=0$ is called a triple collision,
since three particles come together at one point.
The absence of triple collisions in the CH equation is a 
nontrivial result \cite{bss-moment,holden-raynaud-mp},
but for the linearly forced Burgers equation it is much simpler.
(Note, however, that triple collisions \emph{are} possible for discontinuous
piecewise linear solution; see the examples at the end of the paper.)

\begin{theorem}
  Collisions occuring in continuous piecewise linear solutions
  of the linearly forced Burgers equation \eqref{eq:fB}
  cannot be triple collisions.
\end{theorem}

\begin{proof}
  A triple collision would occur if $l_{k-1}(t_0)=0=l_k(t_0)$ for some~$t_0$,
  which amounts to $F_{k-1}(t_0)=0=F_k(t_0)$ by \eqref{eq:ODEsolution-lk}.
  From the definition of $F_k$, it is obvious that this is impossible in the case $M=0$, since we are
  assuming $m_k \ne 0$.
  In the case $M \ne 0$, it is also impossible, although less obvious;
  $F_k(t_0)=0$ iff
  $t_0 = \frac{1}{2M} \log\frac{-\sum_{j>k} m_j(0)}{m_k(0) + \sum_{j<k} m_j(0)}$
  and the quotient inside the logarithm is positive.
  So $F_k=F_{k-1}=0$ iff
  $\frac{m_k(0) + B}{A} = \frac{B}{m_k(0) + A} < 0$,
  where $A = \sum_{j<k} m_j(0)$ and $B = \sum_{j>k} m_j(0)$,
  which requires that $(m_k+A)(m_k+B)=AB$, and hence $m_k(A+m_k+B)=0$.
  But this is ruled out by $m_k$ and $A+m_k+B=M$ both being nonzero.
\end{proof}

We finish this section with the postponed proof of the main theorem.

\begin{proof}[Proof of Theorem~\ref{thm:ODE-solution}]
  Assume to begin with that $x_1(0) < \dots < x_n(0)$.
  Then \eqref{eq:ODEs-identitysector} is equivalent to
  \eqref{eq:ODEs}, and we can attack the problem by trying
  to find $x_k(t)$ and $m_k(t)$ such that
  the corresponding piecewise linear $u(x,t)$ given by \eqref{eq:u-ansatz}
  satisfies the PDE \eqref{eq:fB-with-ansatz}.
  The $x_k$'s divide the real line into $n+1$ intervals which
  we number by $k=0,\dots,n$.
  In each such interval $u$ takes the form $u(x,t)=a_k(t)x+b_k(t)$.
  Inserting this into \eqref{eq:fB-with-ansatz} yields
  $\dot{a}_k + a_k^2 = M^2$ and
  $\dot{b}_k + b_k a_k = -MM_+$,
  from which (in the case $M\ne 0$)
  \begin{equation}
    \label{eq:ak}
    a_k(t) = M
    \frac{a_k(0) \cosh(Mt) + M \sinh(Mt)}{a_k(0) \sinh(Mt) + M \cosh(Mt)}
  \end{equation}
  is found immediately,
  and by making an ansatz for $b_k$ with the same
  denominator as $a_k$ one also obtains
  \begin{equation}
    \label{eq:bk}
    b_k(t) =
    \frac{a_k(0) M_+ \bigl(1-\cosh(Mt) \bigr)
      + M \bigl(b_k(0)- M_+ \sinh(Mt) \bigr)}
    {a_k(0) \sinh(Mt) + M \cosh(Mt)}.
  \end{equation}
  Now $x_k(t)$ and $m_k(t)$ are recovered from the relations
  $m_k = \frac12 (a_k - a_{k-1})$
  and $x_k = -(b_k - b_{k-1}) / (a_k - a_{k-1})$.
  Because of the algebraic nature of the formulas thus obtained,
  they satisfy the ODEs
  \eqref{eq:ODEs-identitysector} identically,
  which shows that the assumption $x_1 < \ldots < x_n$ is
  immaterial and can be removed.
  (This will be important later; see the comments after
  Theorem~\ref{thm:substitution}.)
  The simpler case $M=0$ (unforced Burgers) is entirely similar, except that
  \begin{equation}
    a_k(t) = \frac{a_k(0)}{t a_k(0) + 1},
    \qquad
    b_k(t) = \frac{b_k(0)}{t a_k(0) + 1}.
  \end{equation}
  (The solution for $M=0$ can also be obtained by expanding
  $e^{\pm Mt}=1\pm Mt+O(M^2)$ in the solution for $M\ne 0$ and
  letting $M\to 0$.)
\end{proof}

\section{Inverse spectral construction of solutions}

The Lax pair
\begin{align}
  \label{eq:lax1}
  - \partial_x^3 \phi &= z m\phi,
  \\
  \label{eq:lax2}
  \phi_t &= \left[ z^{-1} \partial_x^2 + c + u_x - u \partial_x \right] \phi,
\end{align}
with $c$ an arbitrary constant,
is compatible iff $m_t+m_x u + 3m u_x=0$ and $m_x=u_{xxx}$, under 
the assumption of sufficient smoothness needed to justify the 
cross-differentiation.
In particular, it is compatible if $u$ evolves according to the derivative
Burgers equation \eqref{eq:dB}, which can be written as
$m_t+m_x u + 3m u_x=0$ with $m=u_{xx}$.
To obtain the linearly forced Burgers equation \eqref{eq:fB} from
equation $\eqref{eq:dB}$ the rule $(u^2)_x = 2uu_x$ is used.
It is not obvious if all these formal calculations have any relevance
to weak solutions, where the smoothness assumptions may be violated.
To investigate this,
let us say that \eqref{eq:lax1} and \eqref{eq:lax2} constitute a
\emph{weak Lax pair} if they are satisfied in the weak sense discussed
in the introduction
(thus $\phi$, like $u$, is a $\mathcal{D}'(\R)$-valued function of $t$,
and the equations hold in the space of distributions $\mathcal{D}'(\R)$).
Solutions $u$ of the form \eqref{eq:u-ansatz},
$u=\sum m_k \abs{x-x_k}$,
do admit a weak Lax pair with $m=u_{xx}=2\sum _{k=1}^n m_k \delta_{x_k}$,
and $\phi$ is in this case a continuous function (in fact, it is piecewise
a quadratic polynomial in $x$ with $t$-dependent coefficients).
The product $m\phi$ in \eqref{eq:lax1} is well-defined since the distribution~$m$
can be multiplied by the continuous function~$\phi$.
We hope to treat weak Lax pairs in more depth in future papers.
Here we just state a theorem
which can be verified by careful use of the calculus of distributions.

\begin{theorem}
  The following are equivalent conditions on a function $u$ of the form
  \eqref{eq:u-ansatz}, $u=\sum m_k \abs{x-x_k}$:
  \begin{enumerate}
  \item $u$ is a weak solution to the linearly forced Burgers equation \eqref{eq:fB},
    and $\{x_k, m_k\}$ satisfy equations \eqref{eq:ODEs}.
  \item $u$ has a weak Lax pair \eqref{eq:lax1}, \eqref{eq:lax2}.
  \end{enumerate}
\end{theorem}

When $u=\sum m_k \abs{x-x_k}$, a solution
to equation \eqref{eq:lax1} with the asymptotic condition
$\phi(x,t;z)=1$ for $x < x_1(t)$ will be consistent with the time evolution
given by \eqref{eq:lax2} provided that we choose the constant $c=-M$.
Such a solution evaluated at
$x > x_n(t)$ will take the form
$\phi(x,t;z)=A(t;z) \frac12 (x-x_n)^2 + B(t;z)(x-x_n) + C(t;z)$,
where all three coefficients are polynomials in $z$, which,
by equation \eqref{eq:lax2}, satisfy
$\dot{A}=0$, $\dot{B}=MB$, and $\dot{C}=\frac{A}{z}+2MC$
(see \cite{kls}).
Thus it is consistent with equations \eqref{eq:lax1} and \eqref{eq:lax2} 
to impose the condition $A(t;z)=0$, which together with
$\phi=1$ for $x<x_1$ amounts to the boundary conditions
$\phi_{x}(-\infty)=\phi_{xx}(-\infty)=\phi_{xx}(\infty)=0$.
With these boundary conditions in place, the problem of solving
the ODEs \eqref{eq:ODEs} becomes an isospectral deformation problem
which can be solved 
if one knows how to solve the inverse problem for equation \eqref{eq:lax1}.  
This is exactly the inverse problem that was studied in \cite{kls} under the
additional assumption that all $m_k(0)>0$.
We now give a brief summary of results from that paper.

\begin{theorem}
  The ``Neumann-like discrete cubic string''
  boundary value problem 
  \begin{equation*}
    - \partial_x^3 \phi= z m\phi,
    \qquad \phi_{x}(-\infty)=\phi_{xx}(-\infty)=\phi_{xx}(\infty)=0, 
  \end{equation*}
  where $m = 2\sum_{k=1}^n m_k \delta_{x_k}$ with all $m_k>0$,
  has a spectrum of the form $\{0=z_0<z_1<z_2<\cdots< z_{n-1}\}$.
  There is a one-to-one (up to translations of $m$ along the $x$ axis)
  and onto spectral map
  $m \mapsto \{M, \mu \}$,
  where $M=\sum m_k > 0$ and $\mu$ is a measure of the form
  $\mu = \sum_{j=1}^{n-1} b_j \delta_{z_j}$,
  with $b_j>0$ for $j=1,\dots,n-1$
  (see details in \cite{kls}).
  The inverse problem of recovering the discrete measure $m$ from $\{M, \mu \}$
  has the explicit solution
  \begin{equation}
    \label{eq:l-m}
    m_{n-k}=\frac{\mathcal{C}_k \mathcal{D}_k}{2\mathcal{A}_{k+1}\mathcal{A}_k},
    \qquad
    x_{n-k+1} - x_{n-k} \equiv l_{n-k} = -\frac{2 \mathcal{A}_k}{\mathcal{D}'_k}.
  \end{equation}
  in terms of determinants of bimoment matrices constructed out of the
  measure $\mu$ and the constant $M$ (see below).
\end{theorem}

We recall the following definitions from \cite{kls}. 
Given a measure $\mu$, let
\begin{equation}
  \label{eq:Ibeta}
  \beta_j=\int z^j \, d\mu(z),
  \qquad
  I_{ij} = I_{ji} = \iint \frac{z^i \, w^j}{z+w} d\mu(z) d\mu(w).
\end{equation}
Let $\mathcal{A}_0=\mathcal{B}_0=\mathcal{C}_0=\mathcal{D}_0=1$,
$\mathcal{A}_1=I_{00}+\frac{1}{2M}$, $\mathcal{D}'_1=\beta_0$, and
for other values of $k$ let
\begin{gather}
  \mathcal{A}_k = \begin{vmatrix}
    I_{00}+\frac{1}{2M} & I_{01} & \cdots & I_{0,k-1}\\
    I_{10} & I_{11} & \cdots & I_{1,k-1} \\
    I_{20} & I_{21} & \cdots & I_{2,k-1} \\
    \vdots & \vdots && \vdots \\
    I_{k-1,0} & I_{k-1,1} & \cdots & I_{k-1,k-1}
  \end{vmatrix},
  \nonumber
  \\
  \mathcal{B}_k = \begin{vmatrix}
    I_{00} & I_{01} & \cdots & I_{0,k-1}\\
    I_{10} & I_{11} & \cdots & I_{1,k-1} \\
    \vdots & \vdots && \vdots \\
    I_{k-1,0} & I_{k-1,1} & \cdots & I_{k-1,k-1}
  \end{vmatrix},
  \qquad
  \mathcal{C}_k = \begin{vmatrix}
    I_{11}& I_{12} & \cdots & I_{1k} \\
    I_{21} & I_{22} & \cdots & I_{2k} \\
    \vdots & \vdots && \vdots \\
    I_{k1} & I_{k2} & \cdots & I_{kk}
  \end{vmatrix},
  \\
  \mathcal{D}_k = \begin{vmatrix}
    I_{10} & I_{11} & \cdots & I_{1,k-1} \\
    I_{20} & I_{21} & \cdots & I_{2,k-1} \\
    \vdots & \vdots && \vdots \\
    I_{k0}& I_{k1} & \cdots & I_{k,k-1}
  \end{vmatrix},
  \qquad
  \mathcal{D}'_k=\begin{vmatrix}
    \beta_0& I_{10} & \cdots & I_{1,k-2} \\
    \beta_1 & I_{20} & \cdots & I_{2,k-2} \\
    \vdots & \vdots && \vdots \\
    \beta_{k-1} & I_{k0} & \cdots & I_{k,k-2}
  \end{vmatrix}.
  \nonumber
\end{gather}
In all these cases,
the index $k$ agrees with the size $k\times k$ of the determinant.
Note that
$\mathcal{A}_k = \mathcal{B}_k + \frac{1}{2M} \mathcal{C}_{k-1}$
for $k\ge 1$.

Let us analyze the formula \eqref{eq:l-m} for $l_k$
in order to compare it with \eqref{eq:ODEsolution-lk}
obtained earlier. First, \eqref{eq:lax2} implies that
the linearly forced Burgers equation induces a very simple
evolution of the measure $\mu$, namely
$\mu(z;t)=e^{Mt} \mu(z;0)$.
Because of this it is easy to factor out the
time dependence from all the determinants involved in \eqref{eq:l-m}.
This elementary exercise leads to $l_k(t)=l_k(0) F_k(t)$, where
\begin{equation}
  F_k(t) =
  \frac{\mathcal{B}_{n-k}(0) \, e^{Mt}
    + \frac{1}{2M} C_{n-k-1}(0) \, e^{-Mt}}
  {\mathcal{B}_{n-k}(0) + \frac{1}{2M}C_{n-k-1}(0)}.
\end{equation}
This is in full agreement with \eqref{eq:Fk} and \eqref{eq:ODEsolution-lk}.
The formula for $m_k$ can be checked in a similar way.

\section{Discontinuous piecewise linear solutions}

\begin{theorem}
  \label{thm:shock-ODEs}
  The discontinuous piecewise linear ansatz \eqref{eq:u-shock-ansatz},
  $u=\sum (m_k \abs{x-x_k} - s_k \sgn(x-x_k))$,
  is a weak solution of the linearly forced inviscid Burgers equation \eqref{eq:fB}
  if and only if
  \begin{equation}
    \label{eq:shock-ODEs}
    \begin{split}
      \dot{x}_k &=\sum_{i=1}^n \bigl( m_i \abs{ x_k-x_i} + s_i \sgn(x_i-x_k) \bigr),
      \\
      \dot{m}_k &= 2 m_k \sum_{i=1}^n m_i \sgn(x_i-x_k),
      \qquad
      \dot{s}_k = s_k \sum_{i=1}^n m_i \sgn(x_i-x_k),
    \end{split}
  \end{equation}
  for $k=1,\ldots,n$.
  For this class of solutions, equation \eqref{eq:fB} takes the form
  \begin{equation}
    \label{eq:shock-fB-with-ansatz}
    u_t+\frac{1}{2}(u^2)_x = M^2 x - M(M_+ + S),
  \end{equation}
  with $M = \sum m_k$ and $M_+ = \sum m_k x_k$ as before, and with
  $S=\sum s_k$.
  The quantities $M$ and $M_+ + S$ are constants of motion,
  and so is $s_k^2 / m_k$ for $k=1,\dots,n$
  (provided that $m_k\ne 0$).
\end{theorem}

\begin{proof}
  We will repeatedly use the following distributional formula valid
  for an arbitrary piecewise differentiable function $f$ with points
  of discontinuity at $x_1,x_2,\ldots, x_n$:
  $f_x = \{f_x\}+ \sum_{k=1}^n [f]_k \delta_{x_k}$,
  where $\{f_x\}$ means the ordinary derivative taken away from
  discontinuities and $[f]_k = f(x_k^+) - f(x_k^-)$
  denotes the jump at $x_k$.
  Moreover,
  $u_t =
  \sum_k \big( \dot{m}_k|x-x_k| -(m_k\dot{x}_k+\dot{s_k})\sgn(x-x_k)
  +2s_k\dot{x}_k \delta_{x_k}\big ).$
  Now the left-hand side of \eqref{eq:fB}, $u_t+\frac12 (u^2)_x$,
  must be a function since the right-hand side is a function;
  hence all Dirac deltas must cancel out.
  Similarly, there must be no Dirac deltas in the first or second $x$ derivatives
  of $u_t+\frac12 (u^2)_x$,
  These conditions give, in turn,
  \begin{equation}
    \begin{split}
      0 &= 2s_k\dot{x}_k + {\textstyle\frac12} [u^2]_k,
      \qquad
      0 = -2(m_k\dot{x}_k+\dot{s_k})+ {\textstyle\frac12} [\{(u^2)_x\}]_k,
      \\
      0 &= 2\dot{m}_k+{\textstyle\frac12} [\{ (u^2)_{xx} \}]_k.
    \end{split}
  \end{equation}
  An elementary computation of jumps for the case of piecewise
  continuous functions now produces \eqref{eq:shock-ODEs}.
  The coefficients of the forcing term in the PDE are identified
  from the smooth part of the term
  $\frac{1}{2}(u^2)_x$, while the constants of motion follow
  from \eqref{eq:shock-ODEs}.
\end{proof}

Weak solutions to an initial value problem are usually not unique unless the
PDE is supplemented with a so-called entropy condition that picks out
the ``physical'' solution. In the case
of the Burgers equation this condition requires $u$ to jump down,
not up, at discontinuities.
This is satisfied by the ansatz \eqref{eq:u-shock-ansatz} if all
shock strenght $s_k$ are nonnegative, so we will assume $s_k \ge 0$
from now on.

When considering the initial value problem for the ODEs \eqref{eq:shock-ODEs}
we can assume without loss of generality that
$x_1(0)<x_2(0)<\ldots<x_n(0)$.  
Thus on a sufficiently small time interval we will still have 
$x_1(t)<x_2(t)<\ldots<x_n(t)$;
in other words, the positions stay in the sector $X_e$
(see \eqref{eq:identitysector}).
In $X_e$ the equations \eqref{eq:shock-ODEs} can be written as
\begin{equation}
  \label{eq:shock-ODEs-identitysector}
  \begin{split}
    \dot{x}_k &=\sum_{i=1}^n \bigl( m_i \sgn(k-i)(x_k-x_i)- s_i \sgn(k-i) \bigr),
    \\
    \dot{m}_k &= 2 m_k \sum_{i=1}^n m_i \sgn(i-k),
    \qquad
    \dot{s}_k = s_k \sum_{i=1}^n m_i \sgn(i-k),
  \end{split}
\end{equation}
for $k=1,\ldots,n$.
These equations can be solved explicitly,
since the simple change of variables in the following theorem reduces
them to the ODEs already solved in Theorem~\ref{thm:ODE-solution}.

\begin{theorem}
  \label{thm:substitution}
  If $\{ x_k, m_k, s_k \}_{k=1}^n$ satisfy
  \eqref{eq:shock-ODEs-identitysector},
  if all $m_k(0)\neq 0$,
  and if
  \begin{equation}
    \label{eq:yk}
    y_k = x_k + s_k/m_k,
  \end{equation}
  then $\{ y_k, m_k \}_{k=1}^n$ satisfy \eqref{eq:ODEs-identitysector}
  (with $y_k$ taking the place of $x_k$ everywhere).
\end{theorem}

\begin{proof}
  Straightforward calculation.
\end{proof}

Note that the initital values $y_k(0)$ will not necessarily be distinct or sorted in
increasing order even though the $x_k(0)$'s are,
but this does not matter since the solution formulas of
Theorem~\ref{thm:ODE-solution} are valid for any initial conditions.
So Theorem~\ref{thm:ODE-solution} gives us $y_k(t)$ and $m_k(t)$
(note that
$\sum m_k y_k = \sum (m_k x_k + s_k) = M_+ + S$
replaces $M_+$ in the solution formula \eqref{eq:ODEsolution}),
and we can then recover $s_k(t)$ from the fact that $s_k^2/m_k$
is constant for each~$k$;
this gives $s_k(t) = s_k(0) / \sqrt{F_{k-1}(t) F_k(t)}$,
and allows us to also recover $x_k(t) = y_k(t) - s_k(t)/m_k(t)$.
This solution $\left\{ x_k, m_k, s_k \right\}$
to \eqref{eq:shock-ODEs-identitysector}
is also the solution to \eqref{eq:shock-ODEs},
at least locally in some time interval around $t=0$
(so that the $x_k$'s remain in the sector $X_e$).

For illustration, here is the general solution with shocks in the case $n=2$,
when $M\neq 0$, $m_1(0)\neq 0$, $m_2(0)\neq 0$:
\begin{equation}
  \label{eq:general-solution-n2}
  \begin{split}
    m_1(t) &= \frac{m_1(0)}{F_0(t) F_1(t)},
    \qquad
    s_1(t) = \frac{s_1(0)}{\sqrt{F_0(t) F_1(t)}},
    \\
    m_2(t) &= \frac{m_2(0)}{F_1(t) F_2(t)},
    \qquad
    s_2(t) = \frac{s_2(0)}{\sqrt{F_1(t) F_2(t)}},
    \\
    x_1(t) &= \frac{M_+ + S - K m_2(0) \, e^{-Mt}}{M} - \frac{s_1(0)}{m_1(0)} \sqrt{F_0(t) F_1(t)},
    \\
    x_2(t) &= \frac{M_+ + S + K m_1(0) \, e^{Mt}}{M} - \frac{s_2(0)}{m_2(0)} \sqrt{F_1(t) F_2(t)},
    \\
    F_0(t) &= e^{-Mt},
    \qquad
    F_1(t) = \frac{m_1(0) \, e^{Mt} + m_2(0) \, e^{-Mt}}{M},
    \qquad
    F_2(t) = e^{Mt},
    \\
    K &= x_2(0) - x_1(0) + \frac{s_2(0)}{m_2(0)} - \frac{s_1(0)}{m_1(0)}.
  \end{split}
\end{equation}

In the continuous case \eqref{eq:ODEs} we assumed all $m_k\ne 0$,
but for \eqref{eq:shock-ODEs} it does make sense to have $m_k=0$
provided that the corresponding $s_k$ is nonzero.
If $m_k(0)=0$, then clearly $m_k(t)=0$ for all~$t$, and the above solution
procedure does not work. But this is easily fixed: just write down the general
solution obtained for a nonzero initial value $m_k(0)=a$,
and let $a\to 0$ there.

We will finish with a few examples that show how to deal with the
solution when it hits the boundary of the sector $X_e$.

\begin{example}
  A particular antisymmetric solution of \eqref{eq:shock-ODEs} with $n=3$ is given by
  $-x_1=x_3 \equiv \xi>0$, $x_2=0$, $-m_1=m_3 \equiv \mu>0$, $m_2=0$, $s_1=s_3=0$,
  $s_2 \equiv \sigma \ge 0$, where
  $\xi(t) = \xi(0) F(t) - \sigma(0) t$,
  $\mu(t) = \mu(0) / F(t)$,
  $\sigma(t) = \sigma(0) / F(t)$,
  with $F(t) = 1 - 2 \mu(0) t$.
  (These formulas are obtained either
  by reducing \eqref{eq:shock-ODEs} to ODEs for $\xi$, $\mu$, $\sigma$
  and solving them directly;
  or by assuming $m_2(0)=a\ne 0$,
  changing variables to $y_1=x_1$, $y_2=x_2+s_2/m_2$, $y_3=x_3$,
  writing down the general solution using Theorems \ref{thm:substitution}
  and~\ref{thm:ODE-solution},
  and letting $a\to 0$;
  or simply by noting that $M=0$ so that we are dealing with the unforced Burgers equation
  whose solution can be found in the textbook way using characteristics.)
  Since $M_+ + S = 2\mu\xi+\sigma$ is constant in time, the wave profile
  (see Figure~\ref{fig:1}) is
  \begin{equation}
    \label{eq:u-antisymm}
    \begin{split}
      u(x,t) &= -\mu(t) \abs{x+\xi(t)} + \mu(t) \abs{x-\xi(t)} - \sigma(t) \sgn(x)
      \\
      &=
      \begin{cases}
        2 \mu(0) \xi(0) + \sigma(0), & x < -\xi(t),\\
        -2 \mu(t) x + \sigma(t), & -\xi(t) \le x < 0,\\
        0, & x=0,\\
        -2 \mu(t) x - \sigma(t), & 0 < x \le \xi(t),\\
        -\bigl(2 \mu(0) \xi(0) + \sigma(0) \bigr), & \xi(t) < x.
      \end{cases}
    \end{split}
  \end{equation}
  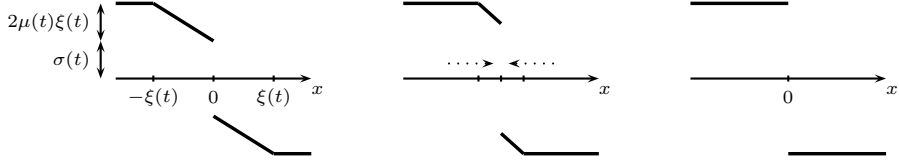
\begin{figure}
    \centering


\begin{pspicture}(-1.5,-1)(1.5,1.5)
  \psline{->}(-1.3,0)(1.3,0) \rput[l](1.3,-.15){\scriptsize $x$}
  \psline[linewidth=1.3pt](-1.3,1)(-0.8,1)
  \psline[linewidth=1.3pt](-0.8,1)(0,0.5)
  \psline[linewidth=1.3pt](0,-0.5)(0.8,-1)
  \psline[linewidth=1.3pt](0.8,-1)(1.3,-1)
  \psline(-0.8,-0.04)(-0.8,0.04) \rput(-0.8,-0.25){\scriptsize $-\xi(t)$}
  \psline( 0.8,-0.04)( 0.8,0.04) \rput( 0.8,-0.25){\scriptsize $ \xi(t)$}
  \psline( 0  ,-0.04)( 0  ,0.04) \rput( 0  ,-0.25){\scriptsize $0$}
  \psline{<->}(-1.5,0)(-1.5,0.5)
  \rput[r](-1.5,0.25){\scriptsize $\sigma(t) \,\,\,$}
  \psline{<->}(-1.5,0.5)(-1.5,1)
  \rput[r](-1.5,0.75){\scriptsize $2\mu(t)\xi(t) \,\,\,$}
\end{pspicture}
\qquad
%
\begin{pspicture}(-1.5,-1)(1.5,1)
  \psline{->}(-1.3,0)(1.3,0) \rput[l](1.3,-.15){\scriptsize $x$}
  \psline[linewidth=1.3pt](-1.3,1)(-0.3,1)
  \psline[linewidth=1.3pt](-0.3,1)(0,0.73)
  \psline[linewidth=1.3pt](0,-0.73)(0.3,-1)
  \psline[linewidth=1.3pt](0.3,-1)(1.3,-1)
  \psline(-0.3,-0.04)(-0.3,0.04)
  \psline( 0.3,-0.04)( 0.3,0.04)
  \psline( 0  ,-0.04)( 0  ,0.04)
  \psline[linestyle=dotted]{->}(-0.8,0.2)(-0.1,0.2)
  \psline[linestyle=dotted]{->}( 0.8,0.2)( 0.1,0.2)
\end{pspicture}
\qquad
\begin{pspicture}(-1.5,-1)(1.5,1)
  \psline{->}(-1.3,0)(1.3,0) \rput[l](1.3,-.15){\scriptsize $x$}
  \psline[linewidth=1.3pt](-1.3,1)(0,1)
  \psline[linewidth=1.3pt](0,-1)(1.3,-1)
  \psline(0,-0.04)(0,0.04) \rput(0,-0.25){\scriptsize $0$}
\end{pspicture}

    \caption{Left/middle: Wave profile $u(x,t)$ as given by
      \eqref{eq:u-antisymm} at two different times $t < \tcoll$, with $\xi(t)$
      decreasing towards zero at a constant rate. Right: Stationary
      profile after collision ($t \ge \tcoll$).}
    \label{fig:1}
  \end{figure}
  If $\sigma(0)=0$ then this is a shockless solution
  (with $n=2$ really, since there is neither mass nor shock at
  the site $x_2=0$).
  It is defined until $\xi(t)=\xi(0) F(t)$ becomes zero at time
  $\tcoll=\bigl( 2\mu(0) \bigr)^{-1}$.
  Then $x_1$ and $x_3$ collide at $x=0$ while $m_1$ and $m_3$ blow up
  to $-\infty$ and $+\infty$, respectively.
  However, $u$ remains bounded, and tends to a shock profile:
  $u(x,t) \to -2 \mu(0) \xi(0) \sgn(x)$
  as $t \nearrow \tcoll$.
  This illustrates that shocks can form naturally even if they are not
  present in the initial wave profile.
  The profile will be stationary after the collision,
  because its continued evolution is given by the $n=1$ case of
  \eqref{eq:shock-ODEs} ($\dot{x}_1 = m_1$, $\dot{m}_1=\dot{s}_1=0$)
  with $x_1=0$, $m_1=0$, $s_1 = 2 \mu(0) \xi(0)$.
  Consequently,
  $u(x,t) = -2 \mu(0) \xi(0) \sgn(x)$ for all $t\ge \tcoll$.

  If $\sigma(0)>0$ there is a shock waiting at the origin between the
  two approaching particles (as in Figure~\ref{fig:1}).
  The solution hits the boundary of the sector $X_e$ when $\xi(t)$ becomes zero
  at time $\tcoll = \bigl( 2\mu(0)+\sigma(0)/\xi(0) \bigr)^{-1}$.
  Then $x_1=x_2=x_3=0$, which illustrates that triple collisions
  may occur when shocks are present.
  Since the collision occurs earlier than in the shockless case,
  $F(t)$ has not yet reached zero at the time of collision;
  hence $m_1$ and $m_3$ do not blow up in this case.
  Again, $u$ tends to a stationary shock profile:
  $u(x,t) = -\bigl(2 \mu(0) \xi(0) + \sigma(0) \bigr) \sgn(x)$
  for all $t\ge \tcoll$.
\end{example}

\begin{example}
  Consider now the shockless ODEs \eqref{eq:ODEs} with $n=3$
  and initial data $m_1(0)=\frac23$, $m_2(0)=-1$ and $m_3(0)=\frac43$,
  so that $M=1$.
  We assume $x_1(0)<x_2(0)<x_3(0)$ but leave them otherwise unspecified.
  Since $u=\pm(Mx-M_+)$ as $x\to\pm\infty$, and since the slope $u_x$ jumps
  by $2m_k$ at each $x_k$, the initial profile $u(x,0)$ consists of line segments
  with slope $-1$, $\frac13$, $-\frac53$ and $1$, joined at the points
  $(x_k, u(x_k,0))$.
  Figure~\ref{fig:2} illustrates this for the particular values
  $x_1(0)=-2$, $x_2(0)=0$, $x_3(0)=1$.
  Note that if the lines $u=\pm(Mx-M_+)$ to the left and to the right
  are continued, they intersect on the $x$ axis at the center of mass
  $x=M_+ / M$ ($=0$ in the figure), which is a constant of motion.

  \begin{figure}
    \centering
    \begin{pspicture}(-5,-0.5)(5,5.5)
  \psline{->}(-5,0)(5,0) \rput[l](5,-.15){\scriptsize $x$}
  \psline[linewidth=1.3pt](-5,5)(-2,2)
  \psline[linewidth=1.3pt](-2,2)(0,2.67)
  \psline[linewidth=1.3pt](0,2.67)(1,1)
  \psline[linewidth=1.3pt](1,1)(5,5)
  \psline(-2,-0.04)(-2,0.04) \rput(-2.3,-0.25){\scriptsize $x_1(0)=-2$}
  \psline( 0,-0.04)( 0,0.04) \rput(-0.3,-0.25){\scriptsize $x_2(0)=0$}
  \psline( 1,-0.04)( 1,0.04) \rput( 1.2,-0.25){\scriptsize $x_3(0)=1$}
  %
  \psline[linewidth=1.3pt, linestyle=dashed](-5,5.04)(-1.5,1.54)
  \psline[linewidth=1.3pt, linestyle=dashed](-1.5,1.5)(0.78,2.78)
  \psline[linewidth=1.3pt, linestyle=dashed](0.78,2.78)(1.33,1.33)
  \psline[linewidth=1.3pt, linestyle=dashed](1.33,1.37)(5,5.04)
  %
  \psline[linewidth=1.3pt, linestyle=dotted](-5,5.08)(-1.2,1.28)
  \psline[linewidth=1.3pt, linestyle=dotted](-1.22,1.28)(1.42,3.10)
  \psline[linewidth=1.3pt, linestyle=dotted](1.42,3.10)(1.67,1.75)
  \psline[linewidth=1.3pt, linestyle=dotted](1.67,1.75)(5,5.08)
\end{pspicture}

    \caption{Solid: Continuous initial wave profile $u(x,0)$. Dashed/dotted:
      $u(x,t)$ at times $t=\ln\frac43$ and $t=\ln\frac53$,
      respectively.}
    \label{fig:2}
  \end{figure}
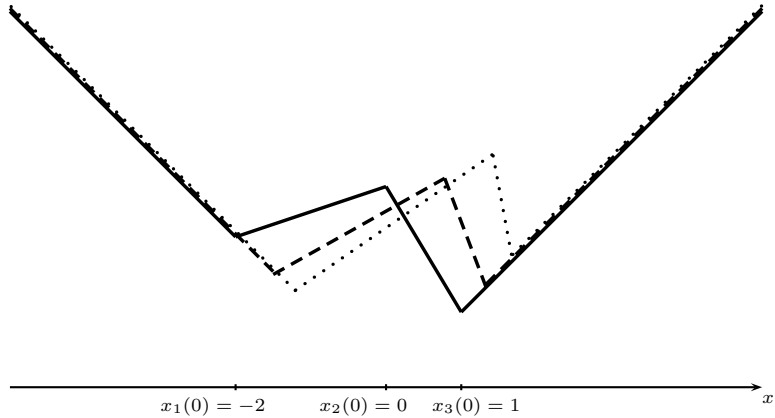

  Recall that $l_1 = x_2-x_1$ and $l_2=x_3-x_2$.
  From \eqref{eq:Fk} we obtain $F_0(t)=e^{-t}$, $F_1(t)=\frac23 e^t + \frac13 e^{-t}$,
  $F_2(t)=-\frac13 e^t + \frac43 e^{-t}$, and $F_3(t)=e^t$.
  Equations \eqref{eq:ODEsolution} and \eqref{eq:ODEsolution-lk} give
  $x_1(t) = x_1(0) + (1-e^{-t}) \bigl( \frac13 l_1(0) + \frac43 l_2(0) \bigr)$,
  $x_2(t) = x_1(t) + l_1(0) F_1(t)$,
  and $x_3(t) = x_2(t) + l_2(0) F_2(t)$.
  There is a collision between $x_2$ and $x_3$ when $F_2(t)$ becomes zero,
  which happens at time $t = \tcoll = \ln 2$ when $e^t=2$.
  At that time we have $F_0=\frac12$, $F_1=\frac32$, $F_2=0$, $F_3=2$,
  hence by \eqref{eq:ODEsolution}
  $m_1=m_1(0)/F_0 F_1=\frac89$, $m_2=-\infty$, $m_3=+\infty$.
  As for the wave profile $u$, we have
  \begin{equation}
    \begin{split}
      u(x_1(t),t)
      &= m_2 l_1 + m_3 (l_1+l_2)
      = (M-m_1) \, l_1 + m_3 l_2
      \\
      &= (F_1 - m_1(0)/F_0) \, l_1(0) + m_3(0) l_2(0) / F_3
      \\
      &\to \textstyle\frac16 \, l_1(0) + \frac23 \, l_2(0),
      \qquad \text{as $t\nearrow \tcoll$},
    \end{split}
  \end{equation}
  and
  \begin{equation}
    \begin{split}
      u(x_2(t),t) - u(x_3(t),t)
      &= (m_1 l_1 + m_3 l_2) - (m_1 (l_1+l_2) + m_2 l_2)
      \\
      &= (m_3 - m_1 - m_2) \, l_2
      \\
      &= \bigl( m_3(0)/F_3 - m_1(0) F_2/F_0 F_1 - m_2(0)/F_1 \bigr) \, l_2(0)
      \\
      &\to \textstyle\frac43 \, l_2(0), \qquad \text{as $t\nearrow \tcoll$}.
    \end{split}
  \end{equation}
  Thus the limiting wave profile at $t=\tcoll$ consists of a line segment
  with slope $-1$, joined to a line segment with slope
  $-1 + 2 \cdot \frac89 = \frac79$ at
  $x=x_1(\tcoll) = x_1(0) + \frac12 \bigl( \frac13 l_1(0) + \frac43 l_2(0) \bigr)$
  and height $u=\frac16 \, l_1(0) + \frac23 \, l_2(0)$;
  the profile jumps down by $\frac43 l_2(0)$ at
  $x=x_2(\tcoll)=x_3(\tcoll)=x_1(\tcoll) + \frac32 l_1(0)$,
  and continues from there with slope $1$.
  See Figure~\ref{fig:3}.

  \begin{figure}
    \centering
    \begin{pspicture}(-5,-0.5)(5,5.5)
  \psline{->}(-5,0)(5,0) \rput[l](5,-.15){\scriptsize $x$}
  \psline[linewidth=1.3pt](-5,5)(-1,1)
  \psline[linewidth=1.3pt](-1,1)(2,3.33)
  \psline[linewidth=1.3pt](2,2)(5,5)
  \psline(-1,-0.04)(-1,0.04) \rput(-1,-0.25){\scriptsize $x_1(\tcoll) = -1$}
  \psline( 2,-0.04)(2,0.04) \rput( 2,-0.25){\scriptsize $x_2(\tcoll)=x_3(\tcoll) = 2$}
  %
  \psline[linewidth=1.3pt, linestyle=dashed](-5,5.04)(-0.61,0.65)
  \psline[linewidth=1.3pt, linestyle=dashed](-0.61,0.61)(3.65,4.49)
  \psline[linewidth=1.3pt, linestyle=dashed](3.65,3.69)(5,5.04)
  %
  %
  \psline[linewidth=1.3pt, linestyle=dotted](-5,5)(0,0)
  \psline[linewidth=1.3pt, linestyle=dotted](0,0)(5,5)
\end{pspicture}

    \caption{Solid: Discontinuous wave profile $u(x,t)$ formed at the
      instant of collision $t=\tcoll=\ln 2$. Dashed: $u(x,t)$ at time
      $t=\tcoll+\frac12$. Dotted: No more collisions occur, and
      $u(x,t) \to \abs{x}$ as $t\to+\infty$.}
    \label{fig:3}
  \end{figure}
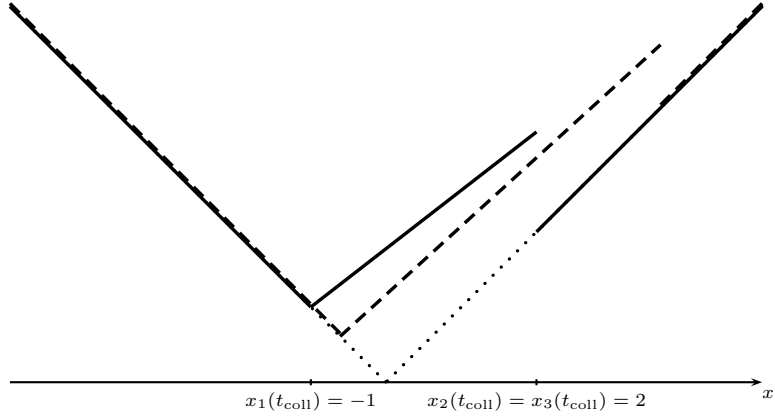

  The continued evolution of the profile for $t\ge \tcoll$
  is illustrated in Figure~\ref{fig:3};
  it is given by the shock ODEs \eqref{eq:shock-ODEs} with $n=2$,
  using a new set of variables whose initial values at $t=\tcoll$ are
  $\tilde{x}_1 = x_1(\tcoll)$, $\tilde{x}_2 = x_2(\tcoll)$,
  $\tilde{m}_1 = \frac89$, $\tilde{m}_2 = \frac19$,
  $\tilde{s}_1 = 0$, and $\tilde{s}_2 = \frac23 l_2(0)$.
  In terms of the new time variable $\tau=t-\tcoll \ge 0$ one finds
  from the general solution \eqref{eq:general-solution-n2} that,
  for example,
  \begin{equation}
    \tilde{x}_2(\tau) - \tilde{x}_1(\tau)
    = \left( \tilde{x}_2(0) - \tilde{x}_1(0)
      + \frac{\tilde{s}_2(0)}{\tilde{m}_2(0)} \right) \tilde{F}_1(\tau)
    - \frac{\tilde{s}_2(0)}{\tilde{m}_2(0)}
    \sqrt{\tilde{F}_1(\tau) \tilde{F}_2(\tau)},
  \end{equation}
  where $\tilde{F}_1(\tau)=\frac89 e^{\tau} + \frac19 e^{-\tau}$
  and $\tilde{F}_2(\tau)=e^{\tau}$.
  Writing this expression as
  $\tilde{x}_2 - \tilde{x}_1 =
  (A+B) \tilde{F}_1 - B \sqrt{\tilde{F}_1 \tilde{F}_2}$,
  we see that it is zero if $F_1(\tau)=0$, which can never happen,
  or if $(A+B)^2 F_1 = B^2 F_2$, which is the same as
  $e^{-2\tau} = 9((A+B)/B)^2-8$ that can't happen either
  since the right-hand side is $>1$ and the left-hand side is $\le 1$
  for $\tau \ge 0$.
  The conclusion is that, in this example,
  $\tilde{x}_2(\tau) - \tilde{x}_1(\tau)$
  remains positive for all $\tau>0$, so there are no more collisions.
  Instead, as $\tau$ (or $t$) $\to+\infty$, we have
  $\tilde{x}_1\to 0$, $\tilde{x}_2\to+\infty$,
  $\tilde{m}_1\to M$, $\tilde{m}_2\to 0$, and $\tilde{s}_2 \to 0$.
  Thus, $u(x,t)$ approaches the limiting wave profile $u(x,+\infty)=\abs{x}$.
\end{example}

\begin{acknowledgments}
  HL's participation in NEEDS 2007 was supported by a travel grant from
  the Knut and Alice Wallenberg Foundation.
  JS is supported by the Natural Sciences and Engineering Research
  Council of Canada (NSERC), Grant No.~138591-04.
\end{acknowledgments}

\bibliographystyle{unsrt}
\bibliography{Lun-Szm-NEEDS2007-arxiv}

\begin{thebibliography}{10}

\bibitem{kls}
Jennifer Kohlenberg, Hans Lundmark, and Jacek Szmigielski.
\newblock The inverse spectral problem for the discrete cubic string.
\newblock {\em Inverse Problems}, 23:99--121, 2007.

\bibitem{dp}
A.~Degasperis and M.~Procesi.
\newblock Asymptotic integrability.
\newblock In A.~Degasperis and G.~Gaeta, editors, {\em Symmetry and
  perturbation theory (Rome, 1998)}, pages 23--37. World Scientific Publishing,
  River Edge, NJ, 1999.

\bibitem{dhh1}
A.~Degasperis, D.~D. Holm, and A.~N.~W. Hone.
\newblock A new integrable equation with peakon solutions.
\newblock {\em Theoretical and Mathematical Physics}, 133:1463--1474, 2002.

\bibitem{ch}
Roberto Camassa and Darryl~D. Holm.
\newblock An integrable shallow water equation with peaked solitons.
\newblock {\em Phys. Rev. Lett.}, 71(11):1661--1664, 1993.

\bibitem{hunter-saxton}
John~K. Hunter and Ralph Saxton.
\newblock Dynamics of director fields.
\newblock {\em SIAM J. Appl. Math.}, 51(6):1498--1521, 1991.

\bibitem{hz1}
John~K. Hunter and Yu~Xi Zheng.
\newblock On a completely integrable nonlinear hyperbolic variational equation.
\newblock {\em Phys. D}, 79(2-4):361--386, 1994.

\bibitem{ls-imrp}
Hans Lundmark and Jacek Szmigielski.
\newblock Degasperis--{P}rocesi peakons and the discrete cubic string.
\newblock {\em IMRP Int. Math. Res. Pap.}, 2005(2):53--116, 2005.

\bibitem{bss-moment}
R.~Beals, D.~Sattinger, and J.~Szmigielski.
\newblock Multipeakons and the classical moment problem.
\newblock {\em Advances in Mathematics}, 154:229--257, 2000.

\bibitem{ls-invprob}
Hans Lundmark and Jacek Szmigielski.
\newblock Multi-peakon solutions of the {D}egasperis--{P}rocesi equation.
\newblock {\em Inverse Problems}, 19:1241--1245, December 2003.

\bibitem{bss-HS}
Richard Beals, David~H. Sattinger, and Jacek Szmigielski.
\newblock Inverse scattering solutions of the {H}unter--{S}axton equation.
\newblock {\em Appl. Anal.}, 78(3-4):255--269, 2001.

\bibitem{bressan-constantin-CH}
Alberto Bressan and Adrian Constantin.
\newblock Global conservative solutions of the {C}amassa--{H}olm equation.
\newblock {\em Arch. Ration. Mech. Anal.}, 183(2):215--239, 2007.

\bibitem{lundmark-shockpeakons}
H.~Lundmark.
\newblock Formation and dynamics of shock waves in the {D}egasperis--{P}rocesi
  equation.
\newblock {\em J. Nonlinear Sci.}, 17(3):169--198, 2007.

\bibitem{holden-raynaud-mp}
Helge Holden and Xavier Raynaud.
\newblock Global conservative multipeakon solutions of the {C}amassa--{H}olm
  equation.
\newblock {\em J. Hyperbolic Differ. Equ.}, 4(1):39--64, 2007.

\end{thebibliography}

\end{document}